\documentclass[twocolumn,showpacs,preprintnumbers,amsmath,amssymb]{revtex4}

\usepackage{graphicx} 
\usepackage{dcolumn} 
\usepackage{bm} 

\begin{document}

\title{Double Au rows on Si(553) surface}

\author{Mariusz Krawiec} 
  \email{krawiec@kft.umcs.lublin.pl}
\affiliation{Institute of Physics, M. Curie-Sk\l odowska University, 
             Pl. M. Curie-Sk\l odowskiej 1, 20-031 Lublin, Poland}

\date{\today}

\begin{abstract}
A new structural model of Au induced Si(553) surface is proposed. The model
accounts for recently experimentally found value of the Au coverage, i.e. 0.48 
monolayer, which suggests formation of two gold chains on each Si(553) terrace. 
The resulting structural model, like the models of other vicinal Si surfaces, 
features the honey-comb chain, but there is no buckling at step edge, which is 
observed on Si(335)-Au and Si(557)-Au surfaces. The present model is more 
stable than the models with single Au chain only, and agrees very well with 
existing experimental data. In particular, calculated band structure, featuring 
two metallic bands coming from hybridization of the gold in both chains with 
neighboring Si atoms, perfectly matches the photoemission data. Moreover, 
theoretical scanning tunneling microscopy topography remains in excellent 
agreement with the experiment. 
\end{abstract}

\maketitle


\section{\label{intro} Introduction}


More than decade ago it has been discovered that submonolayer coverage of gold
stabilizes stepped silicon surfaces, leading to one-dimensional ordering of the 
surface \cite{Jalochowski}. Since then various stepped Si surfaces have been
extensively studied in relation to expected exotic phenomena, characteristic 
for systems of reduced dimensionality \cite{Giamarchi,Himpsel}. Perhaps one of
the most known examples is a gold decorated Si(553) surface. The Si(553)-Au 
surface consists of Si(111) terraces 
$4\frac{1}{3} \times a_{[1 1 \bar{2}]} = 1.48$ nm wide and single atomic steps.
The Si(553) surface normal is tilted from $[111]$ direction towards 
$[1 1 \bar{2}]$ by 12.5$^{\circ}$ \cite{Crain}. The properties of this
surface have been investigated by number of techniques, including scanning
tunneling microscopy (STM)
\cite{Crain,Crain_2,Crain_3,Ahn,Snijders,Crain_4,Crain_5,Ryang,Kang}, angle
resolved photoemission (ARPES) \cite{Crain,Crain_2,Ahn,Crain_5,Barke,Barke_2},
x-ray diffraction \cite{Ghose,Voegeli,Takayama}, and density functional theory
(DFT) \cite{Crain,Riikonen,Riikonen_2,Riikonen_3}. 

Topography of the surface, as measured by STM, features single few nanometers 
long chains on each terrace, which have been identified as originating from the 
step edge Si atoms. Those chains are observed to have 3.84 \AA periodicity 
along the chains, i.e. in $[1 \bar{1} 0]$ direction, which perfectly matches 
the Si lattice constant in this direction. This is completely different from 
what is observed on other vicinal Si surfaces, like Si(335)-Au 
\cite{Crain,MK_1} or Si(557)-Au \cite{Crain,Ahn_2,MK_2}, where chains show 
periodicity doubling, i.e. STM topography shows maxima along the chains with a 
period of $2 \times a_{[1 \bar{1} 0]}$. The periodicity doubling on Si(557)-Au 
and Si(553)-Au surfaces has been explained in terms of the buckling of the step 
edge Si atoms \cite{Riikonen_4,MK_3}, according to which every second step edge
Si atom occupies up (down) position. On Si(553)-Au surface one occasionally 
observes periodicity doubling and even oscillations of topography with a period 
of $3 \times a_{[1 \bar{1} 0]}$ \cite{Snijders}, but this has its origin in 
defects which force the chains to multiply their periodicity \cite{Ryang,Kang}. 
In general, the observed chains on Si(553)-Au surface have the periodicity 
equal to the Si lattice constant in $[1 \bar{1} 0]$ direction. 

The photoemission spectrum of the Si(553)-Au surface is dominated by two
one-dimensional bands (S$_1$ and S$_2$) with parabolic dispersions 
\cite{Crain,Crain_2,Ahn,Barke}. Those bands show different dispersions, and as 
a result cross the Fermi energy E$_F$ at different $k_{\parallel}$, i.e. at 
1.04 (1.07) \AA$^{-1}$, and at 1.18-1.22 (1.25-1.30) \AA$^{-1}$
\cite{Crain_2,Ahn,photovoltaic}. The less dispersive S$_2$ band is split by 85 
meV, and the splitting has been identified to be a spin splitting induced by 
the Rashba spin-orbit (SO) interaction \cite{Barke}. Similar doublet of the 
proximal bands is also observed in Si(557)-Au surface \cite{Losio}, which is 
also spin split due to SO interaction \cite{Sanchez}. The origin of the SO 
split S$_2$ band comes from the hybridization of the row of Au with neighboring 
Si atoms on Si(557) terrace. Therefore it is also expected that, in the case of 
Si(553)-Au surface, situation will be similar. In fact the spin-split S$_2$ 
band has been identified by DFT calculations in one of the models of Ref. 
\cite{Riikonen_3}. However other features of calculated band structure for this 
model disagree with experiment. The other, more dispersive, band (S$_1$) does 
not suffer from the SO interaction and is very similar to the band observed in 
Si(335)-Au surface \cite{Crain}, which also comes from the hybridization of the 
Au row with neighboring Si atoms \cite{MK_4}. The fact, that in case of 
Si(335)-Au surface one does not observe SO splitting (or SO is very small) may 
come from different widths of terraces, and thus interaction of Au row with the 
step edge Si atoms.

Early experimental and theoretical investigations assumed Au coverage to be 
0.24 ML, which is sufficient to form a single row of Au atoms per terrace 
\cite{Crain,Crain_2,Crain_3,Ahn,Snijders,Crain_4,Crain_5,Ryang,Kang,Barke,
Barke_2,Voegeli,Takayama,Riikonen,Riikonen_3}. Thus it is no wonder that DFT
calculations did not reproduce photoemission spectra. Recently determined
Au coverage on Si(553) surface is 0.48 ML, i.e. two Au chains per terrace
\cite{Barke_3}. In fact two Au chains per terrace are consistent with the 
experimentally obtained coverage from x-ray diffraction \cite{Ghose}. However
calculated band structure for the structural model deduced from this experiment 
does not agree with the measured photoemission spectra 
\cite{Riikonen_2,Riikonen_3}. The widely used coverage of single Au chain for 
Si(553) surface traces back to the initial assumption of 2/3 ML coverage for 
Si(111)$\sqrt{3} \times \sqrt{3}$-Au and to 0.44 ML for Si(111)5$\times$2-Au 
reconstructions \cite{Swiech,Bauer}. Thus in light of this new value of Au 
coverage, most of structural models of Si(553)-Au surface need to be revised in 
order to take into account two gold chains per terrace. 

The purpose of the present work is to determine a structural model of the
Si(553)-Au surface, which accommodates two gold chains, and calculate
corresponding band structure. The structural model derived from first 
principles density functional calculations features two Au rows running 
parallel to the step edges and located in the middle of terraces. The step edge 
Si atoms rebond in order to form honey-comb structure, which is also present 
in other vicinal Si surfaces \cite{Crain,MK_4}. This structural model is more 
stable than other structural models with single Au chain 
\cite{Riikonen,Riikonen_3} and the model deduced from x-ray diffraction
\cite{Ghose}. Moreover, the calculated band structure for this model perfectly 
matches the measured ARPES spectra, showing two metallic bands associated with
the hybridizing Au rows with the neighboring Si atoms. The rest of the paper is
organized as follows. In Sec. \ref{details} the details of calculations are
provided. The structural model of Si(553)-Au surface is presented and discussed
in Sec. \ref{structural}, while Secs. \ref{STM} and \ref{band} are devoted to 
the simulated STM topography images and the electronic band structure,
respectively. Finally, Sec. \ref{conclusions} contains some conclusions.


\section{\label{details} Details of calculations}


The calculations have been performed using standard pseudopotential density
functional theory and linear combination of numerical atomic orbitals as a 
basis set, as implemented in the SIESTA code 
\cite{Ordejon,Portal,Artacho,Soler,Artacho_2}. The local density approximation
(LDA) to DFT \cite{Perdew}, and Troullier-Martins norm-conserving
pseudopotentials \cite{Troullier} have been used. In the case of Au 
pseudopotential, the semicore $5d$ states were included. A double-$\zeta$ 
polarized (DZP) basis set was used for all the atomic species 
\cite{Portal,Artacho}. The radii of the orbitals for different species were 
following (in a.u.): Au - 7.20 ($5d$), 6.50 ($6s$) and 5.85 ($6p$), Si - 7.96 
($3s$), 7.98 ($3p$) and 4.49 ($3d$), and H - 7.55 ($1s$) and 2.94 ($2p$).
A Brillouin zone sampling of 12 nonequivalent $k$ points, and a real-space grid 
equivalent to a plane-wave cutoff 100 Ry have been employed. 

The Si(553)-Au system has been modeled by four silicon double layers and a
vacuum region of 19 \AA. All the atomic positions were relaxed until the maximum
force in any direction was less than 0.04 eV/\AA, except the bottom layer. The 
Si atoms in the bottom layer were fixed at their bulk ideal positions and 
saturated with hydrogen. To avoid artificial stresses, the lattice constant of 
Si was fixed at the calculated value, 5.39 \AA.


\section{\label{structural} Structural models}


The total energy calculations of the Si(553)-Au surface show, like in case of 
other Si surfaces \cite{Crain,Riikonen_4,MK_4}, that it is energetically 
favorable for the Au atoms to substitute into the top Si layer. The surface 
energy gain per unit cell is more than 1 eV, as compared to the adsorption of 
Au. Therefore, in the following, I will focus on the structural models of the 
Si(553)-Au surface featuring the top Si layer atoms substituted by the gold.

\subsection{\label{single_cell} Single unit cell}

Since the STM measurements 
\cite{Crain,Crain_2,Crain_3,Ahn,Snijders,Crain_4,Crain_5,Ryang,Kang} show that
the topography modulation along the chains is equal to the Si lattice constant
in direction $[1 \bar 1 0]$, it is natural to take for calculations a single 
unit cell in this direction. This can be also supported by the fact that the
buckling is not observed in Si(553)-Au surface, in contrast to Si(557)-Au and
Si(335)-Au surfaces \cite{Riikonen_4,MK_3}.  

Almost all the proposed structural models of Si(553)-Au surface
\cite{Crain,Riikonen,Riikonen_3} feature single Au chain per terrace, as the
determined Au coverage was twice as small as the actual one \cite{Barke_3}. The 
only model, accounting for two Au chains, was deduced from x-ray diffraction 
experiment \cite{Ghose}, and further investigated by DFT calculations 
\cite{Riikonen_2,Riikonen_3}. According to that model, the Au atoms adsorbed 
near the Si step edges. However, none of the structural models managed to
reproduce the phootemission spectra. Moreover, systematic investigations of 
structural models of the Si(553)-Au surface with proper Au coverage (2 Au atoms 
per Si(553) surface unit cell) show that the more stable models feature the Au 
atoms substituted for the top Si layer atoms in the middle of terrace. 

\begin{figure}[h]                                                              
 \resizebox{0.9\linewidth}{!}{
  \includegraphics{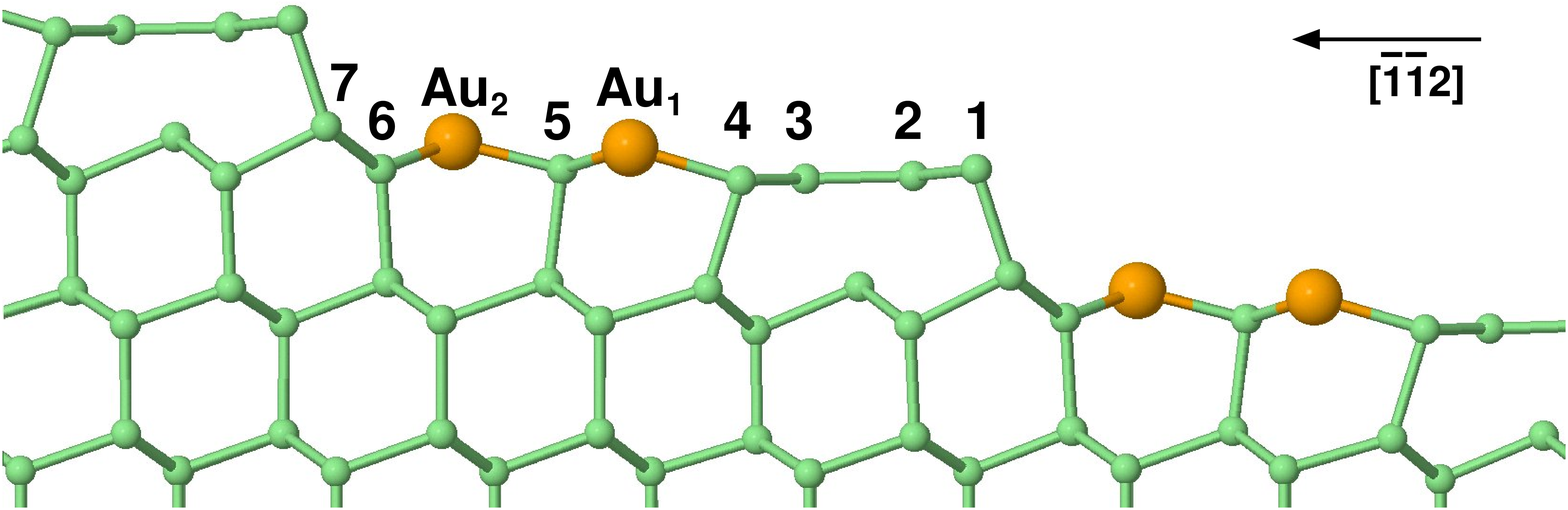}
} \\
 \vspace{0.5cm}
 \resizebox{0.9\linewidth}{!}{
  \includegraphics{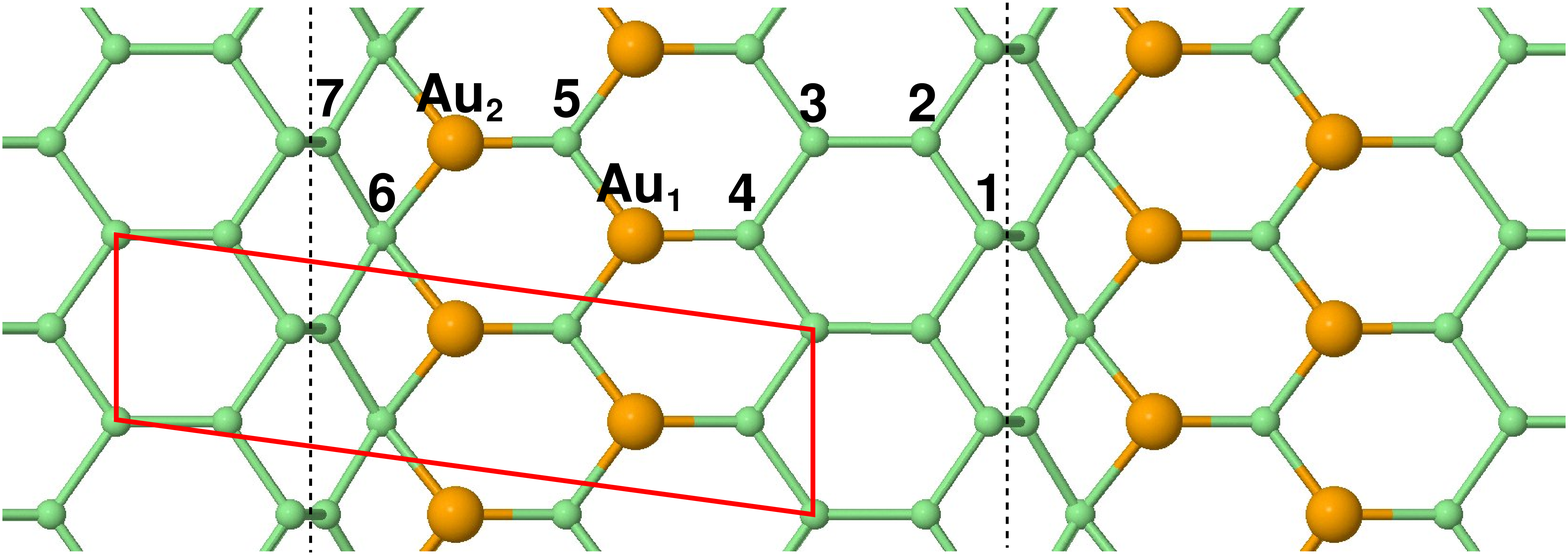}
}
 \caption{\label{Fig1} (Color online) Structural model (single cell model) of 
          the Si(553)-Au surface. Top panel shows side view of the structure, 
	  and bottom panel shows top view with marked surface unit cell. Labels 
	  1-7 stand for silicon surface atoms (Si$_1$-Si$_7$), while Au$_1$ and 
	  Au$_2$ denote gold atoms. The dashed lines in bottom panel indicate 
	  the step edges.}
\end{figure}

Figure \ref{Fig1} shows the most stable structural model (single cell model) of 
the Si(553)-Au surface, where the Si surface atoms (Si$_1$-Si$_7$) are labeled 
by numbers 1-6 and two gold atoms by Au$_1$ and Au$_2$.
This model is energetically more favorable than the most stable Si(553)-Au 
model with single Au row \cite{Crain,Riikonen_4}, and the calculated surface
energy is 16.45 meV/\AA$^2$ lower in the Au-rich limit \cite{chem_pot}. Some of
the other models, in which Au atoms occupy various top layer Si positions, from
Si$_1$ to Si$_7$ (see Fig. \ref{Fig1} for labeling), have slightly higher
energies. The differences are usually in the range of a few meV/\AA$^2$, and the
next 'best' structural model, in which the Au$_2$ occupies the Si$_2$ position,
has energy only 3.48 meV/\AA$^2$ higher. The surface energies of the most 
stable structural models referred to the most stable model with single Au row, 
are summarized in Table \ref{Tab1}.
\begin{table}                                                                 
\caption{\label{Tab1} The relative surface energies of most stable structural 
models of Si(553)-Au structure. The energies are referred to the most stable
model with single Au row.}
\begin{center}                                                                
\begin{tabular}{ccccc}                                                        
\hline                                                                        
& \vline & \\                                                      
position of Au & \vline & surface energy (meV/\AA$^2$)\\
& \vline & \\
\hline
& \vline & \\
Au$_1$, Au$_2$ & \vline & -16.45 \\
Si$_1$, Au$_2$ & \vline & -12.24 \\
Si$_2$, Au$_2$ & \vline & -12.97 \\
Si$_4$, Au$_2$ & \vline & -10.35 \\
Si$_7$, Au$_2$ & \vline & -9.28 \\
& \vline & \\
\hline
\end{tabular}
\end{center}
\end{table}

Although the energy differences between various models listed in Table 
\ref{Tab1} are rather small, there are other arguments supporting the model
shown in Fig. \ref{Fig1}. First one is that the model has the lowest surface
energy. Second one concerns the honey-comb (HC) chain. It is widely accepted 
that the structural models of Au decorated vicinal Si surfaces should posses 
the honey-comb chain at the step edges. The main feature of HC structure is the 
presence of a true double bond between Si$_2$ and Si$_3$ atoms (see Fig. 
\ref{Fig1}), which is responsible for stability of the HC chain \cite{Erwin}. 
However, none of the models, presented in Table \ref{Tab1}, features the HC 
chain at the step edge, except the present best stable model, shown in Fig. 
\ref{Fig1}. Third argument, concerning the band structure, also supports the
present most stable model. The calculated band structure reproduces the
photoemission spectra of Refs. \cite{Crain_2,Crain,Ahn,Barke} reasonably well. 
The other models, which have slightly relative higher surface energies disagree 
with the ARPES spectra. In particular they do not give correct behavior of 
bands near the Fermi energy. This will be discussed in Sec. \ref{band}.

It is known that other Au decorated vicinal Si surfaces, like Si(557)-Au or
Si(335)-Au, show the buckling of the step edge Si atoms \cite{Riikonen_4,MK_3}.
This manifests itself in STM topography images as the periodicity doubling 
along the chains. Since the periodicity of the slab in $[1 \bar 1 0]$ direction 
taken for calculations is equal to the Si lattice constant in this direction, 
all the models discussed above do not take into account the possibility of the 
buckling at step edge, by definition. This is consistent with the experimental
observations 
\cite{Crain,Crain_2,Crain_3,Ahn,Snijders,Crain_4,Crain_5,Ryang,Kang}. However,
it is interesting to check if the doubling of the unit cell in $[1 \bar 1 0]$
direction will leave the present model of the Si(553)-Au surface unchanged.

\subsection{\label{double_cell} Doubling the unit cell}

\begin{figure}[h]                                                              
 \resizebox{0.9\linewidth}{!}{
  \includegraphics{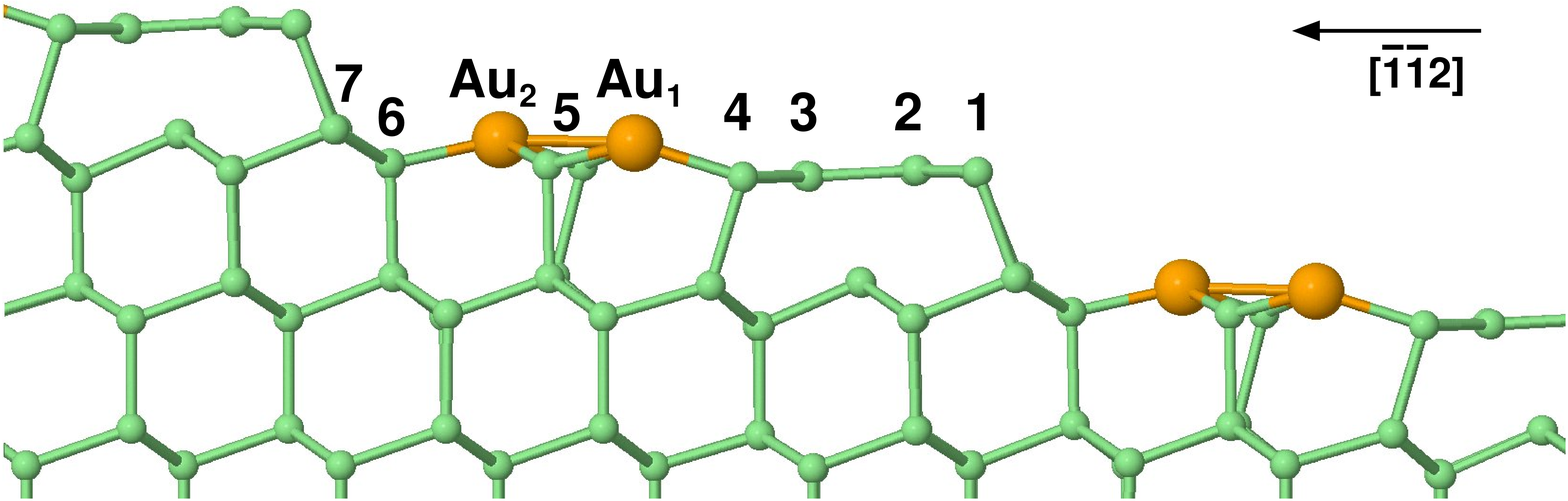}
} \\
 \vspace{0.5cm}
 \resizebox{0.9\linewidth}{!}{
  \includegraphics{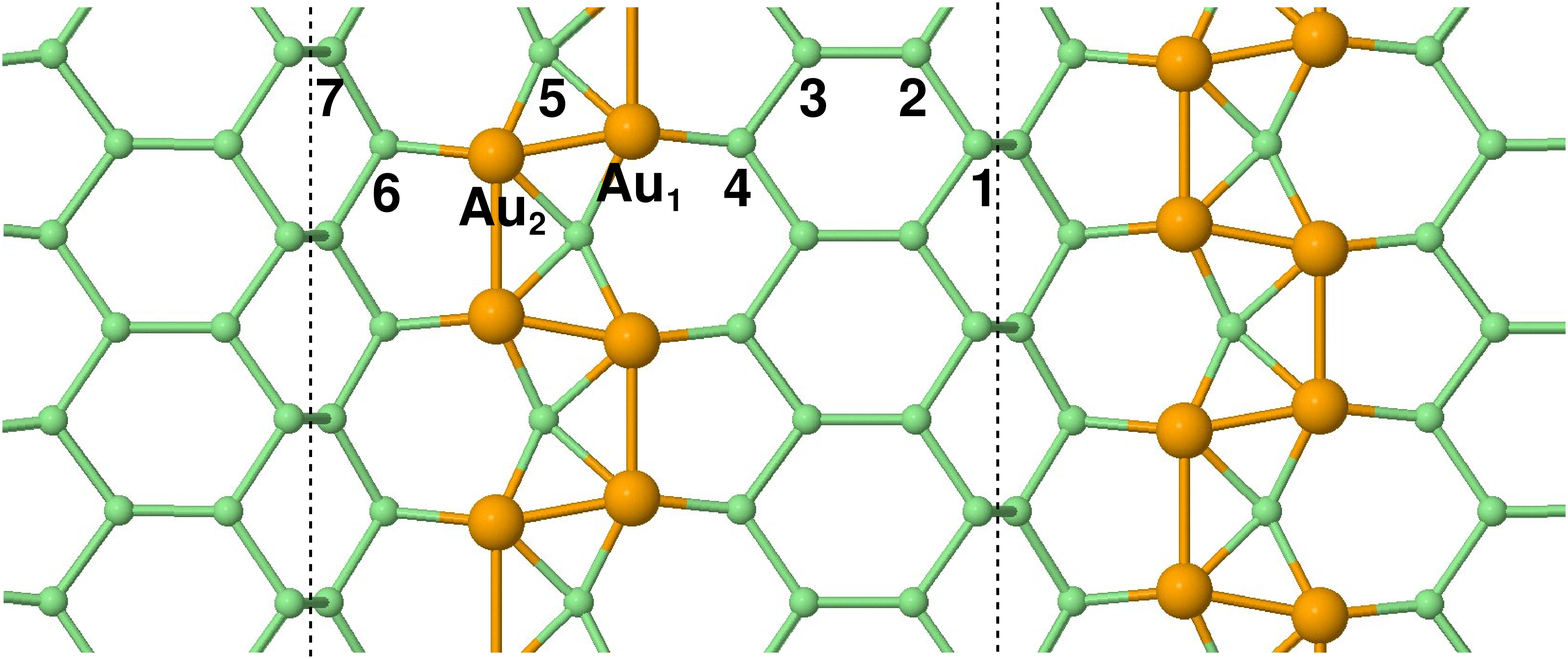}
}
 \caption{\label{Fig2} (Color online) The most stable structural model (double
          cell model) of the Si(553)-Au surface. Top (bottom) panel shows side 
	  (top) view of the structure. Again, labels 1-7 stand for silicon 
	  surface atoms (Si$_1$-Si$_7$), while Au$_1$ and Au$_2$ denote gold 
	  atoms.}
\end{figure}

Doubling the unit cell along the step edges, i.e. in $[1 \bar 1 0]$ direction,
leaves almost all the structural models of the previous subsection unchanged. 
Moreover, it turns out that the models listed in Table \ref{Tab1} have again 
the lowest surface energies. It seems that the Si(553)-Au surface prefers the 
symmetrical arrangement of Au atoms in neighboring surface unit cells. The
models with asymmetrically arranged Au atoms have much higher surface energies.
What is interesting, none of the most stable structural models leads to the
buckling of the step edge Si atoms, in full agreement with the experimental
results \cite{Crain,Crain_2,Crain_3,Ahn,Snijders,Crain_4,Crain_5,Ryang,Kang}. 
Thus one can assume that the single cell model (Fig. \ref{Fig1}), i.e. the 
model with gold atoms in positions Au$_1$ and Au$_2$, is a good candidate for 
structural model of the Si(553)-Au surface. However, a more detailed inspection 
shows that this is not the case. It turns out, that there is a model (double 
cell model), very similar that shown in Fig. \ref{Fig1}, in which gold as well 
as Si$_5$ atoms slightly change their positions, leading to dimerization of Au 
atoms in the rows. The corresponding model is shown in Fig. \ref{Fig2}.
The dimerization of the Au atoms further lowers the surface energy. The energy
gain is 13.73 meV/\AA$^2$, as compared to the single cell model. It is 
worthwhile to note that the local arrangement of the Au and Si$_5$ atoms is the 
same as in recently proposed models of the Si(111)5$\times$2-Au reconstruction 
\cite{Riikonen_5,Ren,Stepniak,Erwin_2}. Moreover, the dimerization of the Au 
atoms leaves the structure at the step edges unchanged, i.e. there is no 
buckling at the step edges, so the periodicity along the terraces still is 
equal to the Si lattice constant in $[1 \bar 1 0]$ direction. However, one can 
notice a sort of horizontal buckling, i.e. the Si$_5$ atoms alternate between 
left and right positions in the direction perpendicular to the steps (see Fig. 
\ref{Fig2}), with distortion $\Delta y = 0.73$ \AA. This fact is reflected in 
STM topography and band structure, as will be discussed in text sections.


\section{\label{STM} STM topography}

\begin{figure}[h]                                                              
 \resizebox{0.9\linewidth}{!}{
  \includegraphics{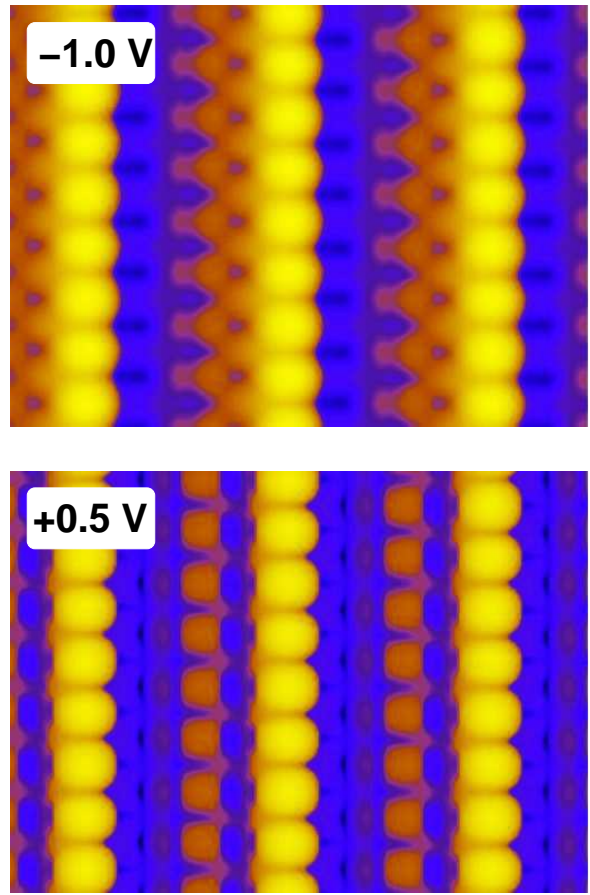}
} 
 \caption{\label{Fig3} (Color online) STM simulations of 4$\times$3 nm$^2$ area
          of the Si(553)-Au surface, calculated for the single cell model. Top
	  (bottom) panel represents filled (empty) state topography,
	  respectively.}
\end{figure}
\begin{figure}[h]                                                              
 \resizebox{0.9\linewidth}{!}{
  \includegraphics{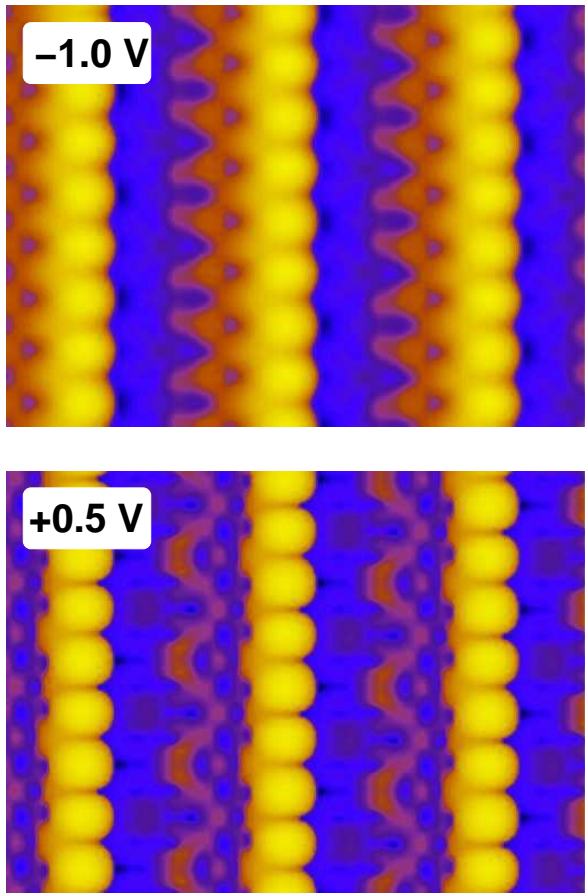}
} 
 \caption{\label{Fig4} (Color online) STM simulations of 4$\times$3 nm$^2$ area
          of filled (top) and empty state (bottom) topography of the Si(553)-Au 
	  surface, calculated for the double cell model.}
\end{figure}

The STM topography data of the Si(553)-Au surface shows one-dimensional
structures, which are interpreted as the step edge Si atoms
\cite{Crain,Riikonen,Riikonen_2,Riikonen_3}. Both structural models discussed 
in Sec. \ref{structural} support this scenario. To further check which one of 
two proposed models, i.e. single cell and double cell models is closer to real 
model of the Si(553)-Au surface, the STM simulations within the Tersoff-Hamann 
approach \cite{Tersoff} have been performed. The results of filled and empty 
state constant current topography calculated for single and double cell models 
are shown in Fig. \ref{Fig3} and in Fig. \ref{Fig4}, respectively.

Figure \ref{Fig3} represents simulated STM topography of 4$\times$3 nm$^2$ of 
the same area of the Si(553)-Au surface for sample bias U=-1.0 V (top panel) 
and U=+0.5 V (bottom panel), obtained within the single cell model.
As it was mentioned previously, the most visible chain structure is associated 
with the step edge Si atoms. The modulation of the topography along these 
chains is equal to the Si lattice constant in $[1 \bar 1 0]$ direction, in 
agreement with STM experiments \cite{Crain_3,Ahn,Crain_4,Ryang,Kang}. Less
visible structure, observed at both polarizations, has been identified as due 
to Si$_4$-Au$_1$ bonds (see Fig. \ref{Fig1}). Similar structure has 
occasionally been observed in STM experiments \cite{Crain_3,Ahn,Ryang}.

The corresponding simulated STM image generated from the double cell model of
the Si(553)-Au surface is shown in Fig. \ref{Fig4}.
Top panel represents filled state (U=-1.0 V), while the bottom one - empty 
state topography (U=+0.5 V). Similar as in the single cell model (Fig. 
\ref{Fig3}), the modulation of topography along the step edge Si chain is equal 
to $a_{[1 \bar{1} 0]}$. As one can notice, the less visible structure in filled
state image (top panel) coming from the bonding of the Si$_4$ and Au$_1$ atoms,
is very similar to that obtained in the single cell model (see top panel of 
Fig. \ref{Fig3}). However, due to the dimerization of the Au$_1$ atoms, the
Si$_4$-Au$_1$ bonds are slightly rotated now, which is reflected in the STM 
image. On the other hand, the empty state topography (bottom panel of Fig.
\ref{Fig4}) is completely different from that of single cell model. We observe 
a sort of zig-zag structure now, which comes from the every second Si$_5$ atom
(closer to the step edge on the same terrace), Au$_1$, Si$_4$ and Si$_3$ atoms
(see Fig. \ref{Fig2}). Moreover, the bonded Si$_5$ and Au$_1$ atoms appear
higher in STM topography image. This could correspond to the experimentally
observed similar structure \cite{Crain_3,Ahn,Crain_4,Ryang,Kang}. 

As we could see, both structural models, giving different topography images,
seem to be consistent with the experimental results. So a question arises, how 
it is possible that we experimentally observe different topography images, once
consistent with the single cell model, and another time, consistent with the
double cell model. To be more precise, the STM topography recorded at the same 
conditions shows areas of different topographies. Some of the surface areas
feature the structure in the middle of terraces with $a_{[1 \bar{1} 0]}$
periodicity, and the other areas show the structures with a periodicity
doubling (see Fig. 1 of Ref. \cite{Crain_3}, Fig. 2 of Ref. \cite{Ahn} or Fig. 
1 of Ref. \cite{Crain_4}). On the other hand, the chain structure associated 
with the step edge Si atoms has always periodicity of the Si lattice constant 
in direction $[1 \bar{1} 0]$. The most plausible scenario for such a behavior 
is that both structures are realized in real Si(553)-Au surface. The structure
obtained within the single cell calculations appears to be a high temperature
phase, while the double cell structure, with lower energy, is a low temperature
phase. At intermediate temperatures, both structures can be locally realized, as
it is evident from the experimental data \cite{Crain_3,Ahn,Crain_4}. Moreover, 
the presence of defects, which could further stabilize one of the phases at 
intermediate temperatures, cannot be omitted. Nevertheless, at very low 
temperature, only the double cell structure is expected to be observable.


\section{\label{band} Band structure}


The experimentally measured electronic band structure features two
one-dimensional bands (S$_1$ and S$_2$) with parabolic dispersions 
\cite{Crain,Crain_2,Ahn,Barke}. Those bands cross the Fermi energy E$_F$ at 
different $k_{\parallel}$, i.e. at 1.07 \AA$^{-1}$ (S$_1$ band), and around 
1.27 \AA$^{-1}$ (S$_2$ band) \cite{Ahn}. The S$_2$ band, crossing the E$_F$ 
near the Si(553) surface Brillouin zone boundary ($k_{\parallel} = 1.27$ 
\AA$^{-1}$ is split by 85 meV, and the splitting has its origin in the Rashba 
spin-orbit interaction \cite{Barke}. Similar doublet of the proximal bands is 
also observed in Si(557)-Au surface \cite{Losio}, which is also spin split due 
to SO interaction \cite{Sanchez}. It is expected that, also in the case of the 
Si(553)-Au surface, situation will be similar. Indeed, the DFT calculations for
one of the models of Ref. \cite{Riikonen_3}, show that S$_2$ band is 
spin-split. However other features of calculated band structure for that model 
disagree with the experimental data. Therefore it is natural to check the band 
structure calculated for the present models against the ARPES data 
\cite{Crain_2,Crain,Ahn,Barke}. Although the SO interaction was not included in
the present calculations, so one cannot expect to get the doublet of proximal
S$_2$ bands, the other features of the band structure should be reproduced 
very well, unless none of the proposed models is correct.

\begin{figure}[h]                                                              
 \resizebox{0.9\linewidth}{!}{
  \includegraphics{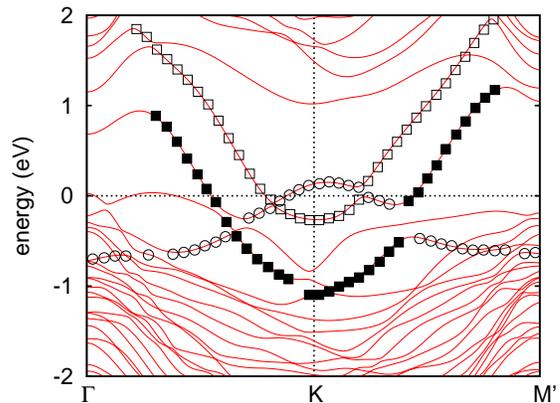}
} 
 \caption{\label{Fig5} (Color online) Band structure of the single cell model,
          calculated in $\Gamma$-K-M' direction of two-dimensional Brillouin
	  zone. The $\Gamma$-K-M' direction is parallel to the steps of the
	  Si(553)-Au surface. The atomic character of the different bands is
	  indicated by using different symbols. Open (filled) squares mark the 
	  band coming from the hybridization of the Au$_1$ and Si$_5$ (Au$_2$
	  and Si$_6$) atoms (see Fig. \ref{Fig1}), while the open circles stand
	  for the step edge Si atoms band. The energies are measured from the
	  Fermi energy (E$_F$ = 0).}
\end{figure}

Figure \ref{Fig5} shows the electronic band structure of the single cell model,
calculated in the direction $[1 \bar 10]$, i.e. parallel to the steps.
A few surface bands are marked with different symbols, reflecting character of
the bands. The band marked with open (filled) squares comes from hybridization
of the Au$_1$ and Si$_5$ (Au$_2$ and Si$_6$) atoms, while the band marked with
open circles comes form the step edge Si atoms. The open and the filled square 
bands are those bands observed in photoemission experiments 
\cite{Crain,Crain_2,Ahn}. Both bands are metallic and cross the Fermi energy at 
k$_{\parallel}$ = 1.00 \AA$^{-1}$ (S$_1$) and 1.18 \AA$^{-1}$ (S$_2$), and 
slightly deviate from the experimentally determined values: 1.04 \AA$^{-1}$ and 
1.18-1.22 \AA$^{-1}$ \cite{Crain_2} or 1.07 \AA$^{-1}$ and 1.25-1.30 \AA$^{-1}$ 
\cite{Ahn}. Moreover, bottom of the S$_1$ band (open squares) appears to be 
shifted towards E$_F$ by 0.4 eV with respect to experimentally determined value 
of -0.64 eV \cite{Crain,Crain_2}. The band marked with open circles, coming 
from the step edge Si atoms is not observed in photoemission, probably due to 
matrix elements. In case of other vicinal Si surfaces, like Si(557)-Au or 
Si(335)-Au, this band is half-occupied \cite{Riikonen_4,MK_3}. This band is 
split due to the buckling of the step edges, and as a result we have two bands: 
one fully occupied, and the other one completely empty. Here, in the case of 
the Si(553)-Au surface, situation is different, as there is no buckling now and 
the band is not split, however due to the presence of additional row of Au 
atoms, this band becomes almost fully occupied.

\begin{figure}[h]                                                              
 \resizebox{0.9\linewidth}{!}{
  \includegraphics{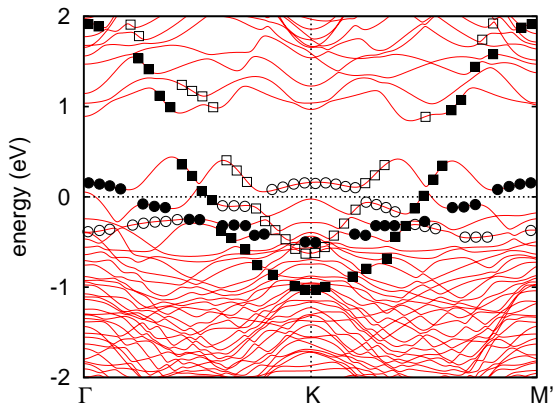}
} 
 \caption{\label{Fig6} (Color online) Band structure of the double cell model,
          calculated in $\Gamma$-K-M' direction of two-dimensional Brillouin
	  zone. Open (filled) squares mark the band coming from the 
	  hybridization of the Au$_1$ and Si$_5$ (Au$_2$ and Si$_6$) atoms (see 
	  Fig. \ref{Fig2}), while the open and filled circles stand for the 
	  step edge Si atoms bands.}
\end{figure}

Slightly different band structure one obtains when calculating it for the 
double cell model, as shown in Fig. \ref{Fig6}.
Again, different symbols stand for different surface bands. Similar as
previously, the band marked with open (filled) squares comes from hybridization
of the Au$_1$ and Si$_5$ (Au$_2$ and Si$_6$) atoms, while the bands marked with
open and filled circles come form the step edge Si atoms. The open and the 
filled square bands are those bands observed in photoemission 
\cite{Crain,Crain_2,Ahn}. They have similar dispersions as in the case of the
single cell model, however the bands cross the Fermi energy at different
k$_{\parallel}$ points, namely at 1.03 \AA$^{-1}$ (S$_1$) and 1.22 \AA$^{-1}$ 
(S$_2$), which are closer to the experimental values, at least those of Ref. 
\cite{Crain_2}. What is also important, energy of the bottom of S$_1$ band 
coincides now with experimentally determined value, i.e. -0.64 eV
\cite{Crain,Crain_2}. The band structure coming from the step edge Si atoms is
now more complicated. Namely, the open circle band of the single cell model is
split now. This splitting has its origin in the dimerization of the Au atoms and
lateral buckling, discussed in Sec. \ref{structural}. The band marked with 
filled (open) circles comes from the step edge Si atoms which are located 
closer (farther) to the Si$_5$ atoms on neighboring terrace. Similar as in case 
of the single cell model, both bands are almost fully occupied owing to the 
presence of the Au atoms.

All the above results show that the double cell model is best candidate for a
structural model of the Si(553)-Au surface, as this model is consistent with 
the STM and ARPES data and features the lowest surface energy. However, it is
also possible that the single cell model is locally realized on the Si(553)-Au
surface.


\section{\label{conclusions} Conclusions}


In conclusion, new structural model of the Au induced Si(553) surface has been 
proposed. The model accounts for experimentally found value of the Au coverage, 
i.e. 0.48 monolayer, which gives two gold chains per Si(553) terrace. Like the
structural models of the other vicinal Si surfaces, the present model features
the honey-comb chain. However, there is no buckling at the step edge. The 
simulated STM topography images show chain structure, associated with the step 
edge Si atoms. The calculated band structure shows two metallic one-dimensional 
bands, coming from the hybridization of Au atoms with neighboring Si atoms. The 
experimentally determined band structure as well as the STM topography are well 
reproduced within the present model.


\begin{acknowledgments}
I would like to thank Prof. M. Ja\l ochowski for valuable discussions. This 
work has been supported by the Polish Ministry of Education and Science under 
Grant No. N N202 1468 33.
\end{acknowledgments}




\begin{thebibliography}{199}
\bibitem{photovoltaic} Differences in k$_{\parallel}$ at which bands cross the
                       Fermi energy come from uncertainty in determination of
		       the E$_F$ due to photovoltaic effect.
\bibitem{chem_pot} The Au chemical potential is taken as the calculated bulk 
                   Au energy per atom.
\end{thebibliography}
\end{document}